\documentclass[%
reprint,
superscriptaddress,
nofootinbib,
amsmath,amssymb,
aps,longbibliography,
pra,
floatfix,
]{revtex4-2}

\usepackage{graphicx}%
\usepackage{verbatim}
\usepackage[version=4]{mhchem}
\usepackage{xcolor}
\usepackage[normalem]{ulem}
\usepackage{pdfpages}
\usepackage{pgffor}
\usepackage{hyperref}
\usepackage{tabularx}
\usepackage{soul}
\usepackage{booktabs}
\usepackage{bm}
\usepackage{multirow}

\makeatletter
\AtBeginDocument{\let\LS@rot\@undefined}
\makeatother

\usepackage[normalem]{ulem}
\definecolor{tangerine}{rgb}{0.944,0.522,0}
\definecolor{verde}{rgb}{0.,0.6,0}
\definecolor{rosso}{rgb}{0.9,0.0,0.2}
\definecolor{orange}{rgb}{1.0,0.5,0.0}

\newif\ifhighlight

\newcommand{\highlight}{\highlighttrue}
\highlight
\newcommand{\editor}[2]{%
  \expandafter\newcommand\csname #1note\endcsname[1]{%
    \textcolor{#2}{(\textbf{#1note:} \textsc{##1})}}%
  \expandafter\newcommand\csname #1\endcsname[1]{%
    \ifhighlight\textcolor{#2}{##1} \else ##1\fi}%
  \expandafter\newcommand\csname #1cancel\endcsname[1]{%
    \ifhighlight\textcolor{#2}{\sout{##1}}\fi}%
  \expandafter\newcommand\csname #1change\endcsname[2]{%
    \ifhighlight\textcolor{#2}{\sout{##1} ##2}\else ##2\fi}%
  \newenvironment{#1text}{\ifhighlight\color{#2}\fi}{\color{black}}
}

\editor{MC}{orange}
\newcommand{\resub}[1]{#1}
\editor{PP}{purple}
\editor{AM}{verde}
\editor{FB}{teal}

\begin{document}

\title{Pushing the limits of unconstrained machine-learned interatomic potentials}

\author{Filippo Bigi}
\email{filippo.bigi@epfl.ch}
\affiliation{Laboratory of Computational Science and Modeling, Institut des Mat\'eriaux, \'Ecole Polytechnique F\'ed\'erale de Lausanne, 1015 Lausanne, Switzerland}

\author{Paolo Pegolo}
\affiliation{Laboratory of Computational Science and Modeling, Institut des Mat\'eriaux, \'Ecole Polytechnique F\'ed\'erale de Lausanne, 1015 Lausanne, Switzerland}

\author{Arslan Mazitov}
\affiliation{Laboratory of Computational Science and Modeling, Institut des Mat\'eriaux, \'Ecole Polytechnique F\'ed\'erale de Lausanne, 1015 Lausanne, Switzerland}

\author{Jonathan Schmidt}
\affiliation{Laboratory of Computational Science and Modeling, Institut des Mat\'eriaux, \'Ecole Polytechnique F\'ed\'erale de Lausanne, 1015 Lausanne, Switzerland}

\author{Michele Ceriotti}
\email{michele.ceriotti@epfl.ch}
\affiliation{Laboratory of Computational Science and Modeling, Institut des Mat\'eriaux, \'Ecole Polytechnique F\'ed\'erale de Lausanne, 1015 Lausanne, Switzerland}

\date{\today}%

\begin{abstract}
Machine-learned interatomic potentials (MLIPs) are increasingly used to replace computationally demanding electronic-structure calculations to model matter at the atomic scale. 
The most commonly used model architectures are constrained to fulfill a number of physical laws exactly, from geometric symmetries to energy conservation. 
Evidence is mounting that relaxing some of these constraints can be beneficial to the efficiency and (somewhat surprisingly) accuracy of MLIPs, even though care should be taken to avoid qualitative failures associated with the breaking of physical symmetries. 
Given the recent trend of \emph{scaling up} models to larger numbers of parameters and training samples, a very important question is how unconstrained MLIPs behave in this limit. 
Here we investigate this issue, showing that -- when trained on large datasets---unconstrained models can be superior in accuracy and speed when compared to physically constrained models. 
We assess these models both in terms of benchmark accuracy and in terms of usability in practical scenarios, focusing on static simulation workflows such as geometry optimization and lattice dynamics.
We conclude that accurate unconstrained models can be applied with confidence, especially since simple inference-time modifications can be used to recover observables that are consistent with the relevant physical symmetries.

\end{abstract}

\maketitle

\newlength{\halfpage}
\setlength{\halfpage}{\columnwidth}

\section{Introduction}

The study of interatomic potentials has long underpinned computational chemistry and materials science, providing a framework for understanding how atoms interact and how their interactions govern stability, reactivity, and thermodynamic behavior. Interatomic potentials are essential in methods such as Monte Carlo simulations~\citep{metr+53jcp}, molecular dynamics~\citep{alde-wain59jcp}, and geometry optimization, where they enable mechanistic exploration of atomic-scale processes across molecular, biological, and condensed-matter systems.
Traditional potentials rely on physically motivated analytic forms, which are cheap to evaluate, but whose accuracy is fundamentally limited by the simplicity of their mathematical formulation.

The past decade has witnessed the widespread adoption of machine-learned interatomic potentials (MLIPs)~\citep{behl-parr07prl}. By training on reference data from quantum-mechanical calculations, MLIPs achieve accuracy approaching first-principles methods at highly reduced computational cost. While initial developments focused on system-specific training, the availability of increasingly diverse datasets~\citep{mptrj-and-chgnet} -- spanning much of the periodic table and containing millions of labeled configurations -- has driven the training of general-purpose (or universal) MLIPs~\citep{m3gnet}, which can afford good-quality predictions across very diverse systems.
This shift has encouraged the adoption of more expressive architectures, notably graph neural networks~\citep{hitchhiker}, as opposed to earlier physically inspired functional forms which might be too rigid or limited~\cite{chong2025resolving}. It has even prompted some models to discard explicit physical symmetries in favor of unconstrained architectures where these are learned from the training data~\citep{gemnet,pozd-ceri23nips,eissler2026}. While unconstrained architectures can achieve high computational efficiency, most often by not enforcing rotational symmetries and/or conservation of energy, models trained for practitioners and popular benchmarks~\citep{matbench} often rely on explicitly enforcing all physical symmetries of the learning target.

In this work, we show that fully unconstrained architectures can be scaled to large and diverse datasets, achieving accuracies on par with state-of-the-art equivariant neural networks. We find that unconstrained models tend to be very efficient at inference time -- a very desirable property in molecular simulations -- although they often require a larger number of epochs to train from scratch, as they need to infer the existence of symmetries and conservation laws from training data.
We also demonstrate their applicability to structural optimization and lattice dynamics, showing that the potential pitfalls of unconstrained models can be corrected with minimal effort.

\section{Background and related work}\label{sec:background-and-related}

\newcommand{\mbf}[1]{\bm{#1}}

\subsection{Interatomic potentials and their properties}

An interatomic potential is a function describing the energy of an atomic structure:
\begin{equation}
V(\{\mbf{r}_i, a_i\}_{i=1}^N),
\end{equation}
where $\mbf{r}_i$ are the three-dimensional positions of the atoms, $a_i$ are their atomic types (chemical elements), and $i$ is an index running over the $N$ atoms in the structure. 

\paragraph{E(3)-invariance.} Interatomic potentials are invariant under the transformations of the Euclidean group in three dimensions E(3), which includes translations, rotations and reflections. Given a group element $g \in E(3)$ acting on all positions, then $V(\{g \cdot \mbf{r}_i, a_i\}_{i=1}^N) = V(\{\mbf{r}_i, a_i\}_{i=1}^N)$.

\paragraph{Permutational invariance.}

Interatomic potentials are invariant with respect to permutations of atom indices, i.e., $V(..,\mbf{r}_i, a_i, .., \mbf{r}_j, a_j, ..) = V(..,\mbf{r}_j, a_j, .., \mbf{r}_i, a_i, ..)$.

\paragraph{Locality.} With few exceptions, atoms that are distant from one another affect the potential energy function independently (in other words, they do not interact~\citep{prod-kohn05pnas}). Mathematically, if atoms $m$ and $n$ are distant (i.e., $|\mbf{r}_m - \mbf{r}_n|$ is large), then $V(\{\mbf{r}_i, a_i\}_{i=1}^N) - V(\{\mbf{r}_i, a_i\}_{i=1,i \neq n}^N) - V(\{\mbf{r}_i, a_i\}_{i=1,i \neq m}^N) + V(\{\mbf{r}_i, a_i\}_{i=1,i \neq m,i \neq n}^N) \approx 0$.\footnote{One can think of it as ``the potential energy change in removing atom $n$ is the same whether atom $m$ is there or not'', i.e. $V(\{\mbf{r}_i, a_i\}_{i=1}^N) - V(\{\mbf{r}_i, a_i\}_{i=1,i \neq n}^N) \approx V(\{\mbf{r}_i, a_i\}_{i=1,i \neq m}^N) - V(\{\mbf{r}_i, a_i\}_{i=1,i \neq m,i \neq n}^N)$. Note that the condition is symmetric in the two atoms $m$ and $n$.}

\subsection{Machine-learned interatomic potentials}

Machine-learned interatomic potentials~\citep{behl-parr07prl,unke+21cr} (MLIPs) address the long-standing trade-off between accuracy and efficiency in atomistic simulations. Classical empirical potentials are computationally cheap, but their simple analytic forms severely limit transferability and predictive power, and they are only available for a few selected systems. Conversely, quantum-mechanical methods such as density functional theory (DFT) provide accurate energies and forces for any system, but their cost restricts applications to small systems or short timescales. MLIPs provide a middle ground: by training flexible function approximators on quantum-mechanical data, they achieve near–first-principles accuracy while maintaining orders-of-magnitude lower computational cost. Graph neural networks (GNNs) have emerged as particularly effective architectures for MLIPs, as they are well-suited to incorporate essential physical symmetries such as permutational and translational invariance, as well as locality by virtue of constructing graph edges based on a cutoff radius. As a result of these aspects and due to their expressivity, GNNs can achieve high accuracy and generalization across chemical systems, explaining their prevalence in current benchmarks~\citep{matbench} and practical applications.

While early MLIPs were tailored to specific systems~\citep{deri+21cr,unke+21cr,behl21cr}, recent years have seen the development of universal MLIPs~\citep{m3gnet,mptrj-and-chgnet,macemp,grace,orbv3,petmad,sevennet,uma} trained on increasingly large and diverse datasets~\cite{mptrj-and-chgnet,alexandria,omat,omol,oc20,oc22,spice,mad}. Universal MLIPs promise to become general-purpose tools for computational materials science and chemistry: pretrained on massive datasets, they can then be applied directly or after inexpensive fine-tuning for downstream tasks. The success of universal MLIPs demonstrates that generalization across diverse systems is attainable and motivates the exploration of increasingly expressive architectures and scaling strategies. In particular, there is growing interest in using functional forms that do not enforce all the physical properties of the potential, but learn them from the data, as will be discussed further in Sec.~\ref{ssec:learnable-symmetries}.

\paragraph{Interatomic potentials and forces}\label{ssec:forces}

Most applications (including molecular dynamics, geometry optimization, phonon calculations; see Sec.~\ref{ssec:applications-and-benchmarks}) make use predominantly or exclusively of interatomic forces, as opposed to potential energies. In this context, forces can be defined as the negative derivative of the potential energy function $V$ with respect to the atomic positions:
\begin{equation}\label{eq:forces}
    \mbf{f}_j = -{\partial V(\{\mbf{r}_i, a_i\}_{i=1}^N)}\mathbin{\big{\slash}}{\partial \mbf{r}_j},
\end{equation}
where $\mbf{f}_j$ denotes the force acting on atom $j$. Although most MLIPs provide forces by automatic differentiation through Eq.~\ref{eq:forces}, several recent state-of-the-art MLIPs are trained and/or provide inference using ``direct" forces~\citep{orbv2,escaip,orbv3,esen,uma}, i.e., forces that are simply predicted as an additional head of the model. 
By avoiding a backward differentiation step, this provides a speedup of a factor between 2 and 3, depending on the architecture~\citep{darkside}, as well as reduced memory usage. Since forces that do not obey Eq.~\ref{eq:forces} are not guaranteed to conserve energy in simulations, they are also referred to as \emph{non-conservative} forces. In contrast, forces obeying Eq.~\ref{eq:forces} are also called \emph{conservative}.

\subsection{Applications and benchmarks}\label{ssec:applications-and-benchmarks}

MLIPs are routinely used for a wide range of applications. Conceptually, the most simple are geometry optimization and lattice dynamics calculations. 
Given a set of atomic types and initial positions, geometry optimization finds a nearby local minimum of the potential energy surface as a function of the atomic positions -- with low-energy minima being candidates for thermodynamically stable configurations. 
A more rigorous assessment of stability can be achieved by a \emph{convex hull} construction, that consists of collecting several local minima, differing by structure or composition, and determining those that have lower (free) energy than a mixture of other phases with the same overall density or chemical composition.
To also incorporate finite-temperature effects one can perform \emph{lattice dynamics} (phonon) calculations, which involve computing the mass-scaled Hessian of the potential around a minimum, and diagonalizing it to obtain its eigenvalues (vibrational frequencies) and eigenvectors (the corresponding atomic displacements). 
From lattice vibrations one can evaluate harmonic free energy corrections to the potential energy of the minima, as well as information on dynamical properties such as infrared or Raman spectra; anharmonic effects can be included considering phonon-phonon interactions via perturbation theory~\citep{peierls1929}.
Alternatively, exact anharmonicity can be obtained by explicitly sampling the Boltzmann distribution at the relevant temperature $T$, i.e. $e^{-V/k_BT}$, which can be achieved through Monte Carlo~\citep{metr+53jcp} or molecular dynamics~\citep{alde-wain59jcp,rahm-stil71jcp,ande80jcp} simulations. 
Sampling based on the potential energy allows the computation of quantities such as order–disorder transition temperatures, phase diagrams, or specific heats. Between the two, molecular dynamics simulations are often preferred for their superior computational efficiency for medium-sized and large systems~\citep{alle-tild17book}. Given that most of these applications rely heavily on interatomic \emph{forces}, rather than only the potential, direct force evaluation can speed up almost every MLIP workflow.

Even though the ultimate test of the utility of a MLIP is whether it can be used reliably for practical atomistic simulation tasks, it is often useful to have standardized benchmarks to encourage the development of more accurate and/or efficient MLIPs. Matbench-discovery~\citep{matbench} was among the first to be introduced, and it features the largest number of state-of-the-art architectures, but it does not contain tests on molecular dynamics applications. Very recently, LAMbench~\citep{lambench} and MLIP Arena~\cite{mlip-arena} have also been proposed. In order to benchmark models for applications in organic chemistry and biochemistry, some practitioners evaluate model accuracy on the various test splits of the SPICE-MACE-OFF dataset~\citep{spice,mace-off}, and the Open Molecules leaderboard and community challenges~\cite{omol} are also increasingly used in this domain.

\section{Theory and methods}\label{sec:theory}

\begin{figure*}[tbp]
    \centering
    \includegraphics[width=0.8\linewidth]{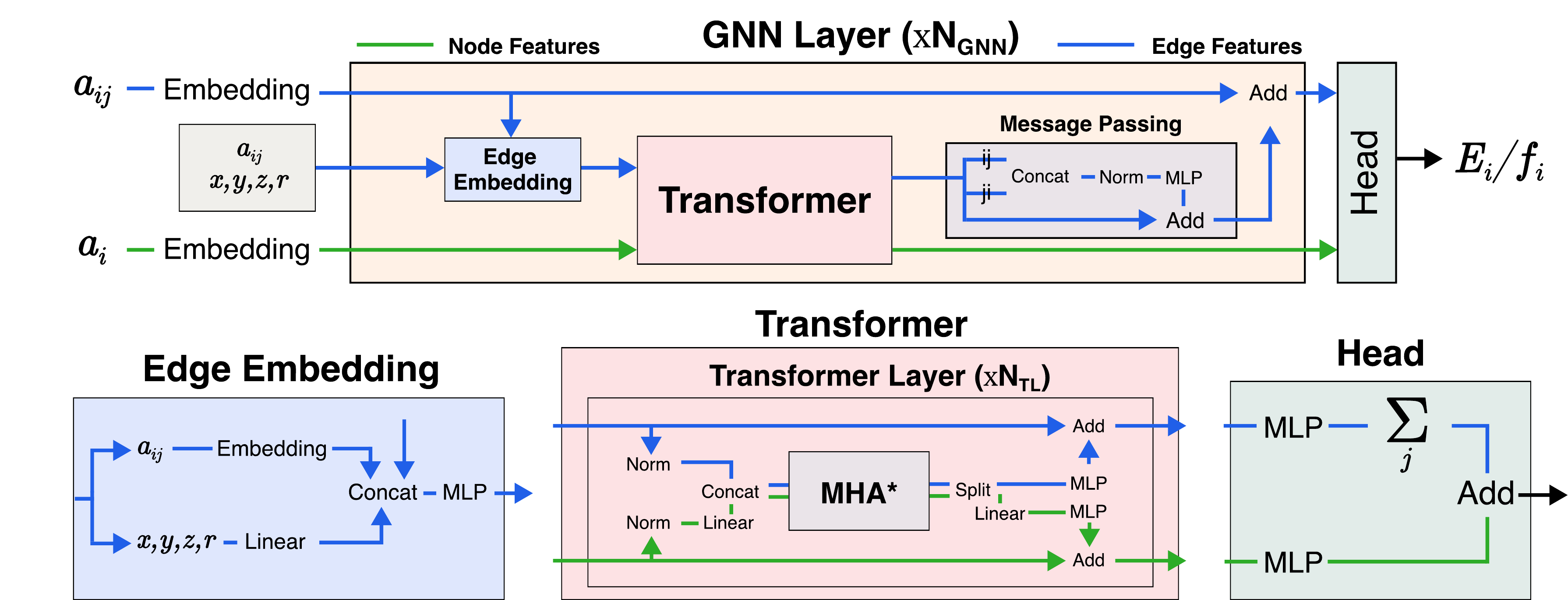}
    \caption{
    Illustration of the proposed architecture. $a_i$ and $a_{ij}$ are the chemical elements of a center atom and a neighbor atom, respectively. $E_i$ represents an atomic energy; all atomic energies are summed to obtain the total energy.\\
    *Attention weights are scaled to ensure smoothness as described in Ref.~\citep{pozd-ceri23nips}.
    }
    \label{fig:pet}
\end{figure*}

The question of whether a physical prior should be built into the functional form of a MLIP or learned as part of the training process is not one with a simple answer. 
The best strategy depends on how constraining the functional form affects the expressivity of the architecture, as well as its computational cost -- which in turn depends on how well the necessary mathematical operations are supported by modern libraries and hardware.
It is also important to consider how hard it is to learn the symmetry, and to monitor and correct for it at inference time. 
Looking at the current landscape, modern GNN-based MLIPs almost universally enforce permutation and translation symmetries, while several recent models relax the requirement for rotational symmetry and energy conservation.

\subsection{What determines whether a symmetry is learnable?}\label{ssec:learnable-symmetries}

One simple and convenient way to learn symmetries is to apply data augmentation during training. The difficulty in learning a symmetry can then be estimated in terms of the number of transformed structures that must be generated to provide symmetry information on a resolution comparable to the scale over which one can expect substantial changes in the learned interatomic potential, which we roughly estimate at about 0.1\AA{} for MLIP applications.

\paragraph{Translations.} Learning the translational invariance of the energy can be considered impossible, as  one would need to perform an infinite number of augmentations. Fully periodic structures, however, require only a finite number of translations within our assumptions: for a 10 \AA~$\times$ 10 \AA~$\times$ 10 \AA~cell, one would need approximately (10 \AA~/~0.1 \AA)$^3$ $\approx 10^6$ augmentations. Besides the fact that this number of augmentations is too large to be practical, the resulting potential energy surface would not be transferable to ranges of Cartesian coordinates outside the training set. In practice, translational invariance is often achieved as a byproduct of enforcing locality, which further explains why it is applied almost universally.

\paragraph{Permutations.} 
Considering a small number of neighbors of 30 within a single atomic environment, one would have to augment each structure $30!$ times to train a permutationally unconstrained model, which is unfeasible. Examples of models using random sampling of permutations appeared in the early days of the field~\citep{mont+12nips}, but modern GNNs almost invariably enforce permutation invariance by making use of permutationally invariant pooling operations.

\paragraph{Rotations.} If we consider the furthest atom that contributes significantly to the description of an atomic environment within a GNN layer to be distant around 4 \AA~from the central atom, a rotationally unconstrained model would have to learn invariance over a spherical surface of $4 \pi (4 \text{\AA})^2$, yielding $4 \pi (4 \text{\AA})^2 / (0.1 \text{\AA})^2 \approx 20\,000$ augmentations, which is attainable, especially for large datasets which contain similar environments with different orientations, and which therefore contribute to the sampling of rotational symmetry. At inference time, it is relatively easy to reduce symmetry breaking by averaging over a grid of rotations, which makes it practical to relax this symmetry constraint.

\paragraph{Inversions.} Data augmentation for the inversion symmetry only involves one additional structure, for a total of two. The low information requirement of this symmetry results in equivariant architectures often attempting to alleviate its cost, either by excluding pseudotensor representation from the neural network \citep{nequip,bata+22nips,allegro}, potentially losing expressivity, or by not enforcing inversion symmetry entirely \citep{equiformer,so3krates,esen}, resulting in SE(3)-invariant predictions as opposed to E(3)-invariant predictions.

\paragraph{Energy conservation.} As far as we are aware, there is no simple way to formulate a conservative constraint in terms of data augmentation (without training on the Hessian of the potential energy function). The energy conservation condition is equivalent to that of a symmetric Jacobian, which is a square matrix of size $3N$, for a total of 8010 constraints for a typical 30-atom training structure. One way to probe this type of symmetry breaking, which could be used as a penalty term during training, consists of checking for the symmetry of the Jacobian of the forces (i.e., the Hessian of the potential for an energy-conserving model). However, this is too expensive to be done in practice, and non-conservative models rely on the presence of geometrically close structures in the dataset to learn approximate energy conservation.

\subsection{Architecture}\label{sub:architecture}

In order to explore the feasibility of training on large datasets in a fully rotationally unconstrained fashion, we choose the PET architecture~\citep{pozd-ceri23nips}, a GNN whose successive layers process the edges within each atomic neighborhood through a standard transformer, where each token corresponds to one edge. To make the architecture and training protocol better adapted to larger-scale datasets, we apply some changes with respect to Ref.~\citep{pozd-ceri23nips}: 
(1) We use a more modern transformer architecture, including root mean square layer normalization~\citep{rmsnorm}, the SwiGLU~\citep{swiglu} activation function, and pre-normalization~\cite{prelayernorm}.
(2) We increase by a factor of four the number of node features for a given edge feature size (the two are instead the same in Ref.~\cite{pozd-ceri23nips}). Indeed, processing node features is extremely cheap compared to edge features, as there are typically 30-50 times more edges than nodes in a chemical structure with typical cutoff radii. This allows one to greatly increase the parameter counts of the models while adding nearly zero overhead to training and inference timings and memory usage. This change is applicable to nearly all GNN-based MLIP architectures, not only unconstrained ones. 
(3) We do not discard node features at every GNN layer, but pass them to the next. This is important in architectures with a larger number of GNN layers (which are used in this work) and where the node features constitute a larger fraction of the total representation power of the neural network. 
(4) We change the learning schedule to a standard linear warm-up followed by cosine decay, as opposed to successive learning rate reductions. (5) Direct forces can be predicted by an additional head that outputs a 3-vector for each atom, similar to how atomic energies are predicted. (6) For some of our models (see App.~\ref{app:model}), we use an adaptive (but smooth and featuring linear time and memory complexity in the number of neighbors) cutoff strategy, described in App.~\ref{app:adaptive-cutoff}. A more comprehensive description of the architecture is available in Fig.~\ref{fig:pet} and App.~\ref{app:model}. 

\subsection{Obtaining physical observables from unconstrained models}

While obtaining physically accurate observables from unconstrained models has been demonstrated in some applications, this practice has not been established in general. In most cases small amounts of symmetry breaking do not quantitatively affect the predicted properties (e.g., the relative stability of different phases or compositions when performing a convex-hull construction).
However, some care is required to avoid qualitative changes in results, especially within automated workflows.
It is worth remembering that the practical implementations of many electronic-structure methods often introduce similar artifacts: for instance, real-space grids break translational and rotational symmetry, an issue that is still actively worked on~\citep{durh+25es}.
For symmetries associated with a compact group, it is always possible to reduce or eliminate the symmetry error by inference-time augmentation, e.g., summing over a Lebedev grid~\citep{lang+24mlst}, or by ensembling~\citep{gerken2024emergentequivariancedeepensembles}.

\paragraph{Molecular dynamics.} Non-equivariant models can often be used out of the box in molecular dynamics simulations with minimal downsides~\citep{lang+24mlst,petmad}. Systematic long-time errors can be avoided by evaluating the MLIP and its forces on a different random rotation (and inversion) of the structure at every time step~\citep{lang+24mlst}, with negligible overhead. %
Direct force models can also recover quantitative physical observables in molecular dynamics~\citep{darkside} using a multiple time-stepping algorithm~\citep{tuck+92jcp} where conservative and non-conservative forces are used together, with the computationally advantageous non-conservative forces constituting the majority of the evaluations.

\paragraph{Geometry optimization.} Geometry optimization with non-conservative force models was investigated in Ref.~\citep{darkside}, showing that, although inaccurate direct-force models struggle to converge, accurate models have nearly no qualitative downsides. This is intuitive, as the target force field is conservative, and therefore the model must be conservative at least up to its force error.
A similar investigation on the effects of using rotationally unconstrained models for geometry optimization (Sec.~\ref{ssec:geop-and-phonons}) shows that symmetry breaking is possible but benign, and possibly even beneficial as it can relax away from unstable high-symmetry structures. It might however be necessary to use more permissive thresholds to identify the symmetry of the relaxed structure than when using exactly equivariant models. 
Whenever the space group is known, it is also possible to restore exact symmetry by projecting out the component of the forces incompatible with the group action, or by averaging the model predictions over the symmetries of the group. If computational cost is not a limiting factor, averaging over inversions and/or a grid of rotations is also possible.

\paragraph{Frequency and phonon calculations.} 

Phonon calculations often assume exact symmetry of the underlying structure, since phonon bands are plotted along high-symmetry lines in the lattice Brillouin zone (BZ). This makes it especially important to either perform a constrained-symmetry optimization, relax the symmetry-detection threshold, or fix the BZ sampling path to that of the high-symmetry group.
The residual stress left after symmetry-constrained relaxation due to the lack of equivariance is sufficiently small that phonons can be computed at the symmetric minimum without artifacts such as spurious soft modes. At the unconstrained, slightly distorted minimum, BZ-integrated quantities like the phonon density of states are practically indistinguishable from the symmetric case, indicating that the absence of exact equivariance has little impact on global properties. 
Direct-force models require additional care, as they do not guarantee a vanishing total force, and their Jacobian with respect to atomic displacements is not the (symmetric) Hessian of a potential energy function. In practice, both conditions are easy to meet: the net force is manually subtracted from the predictions, and the Jacobian is symmetrized by default in most phonon codes (see Sec.~\ref{ssec:geop-and-phonons}).

\section{Results}
\newcommand{\B}[1]{\textbf{#1}}
\newcommand{\U}[1]{\underline{#1}}

\subsection{Large-scale materials datasets}\label{ssec:large-scale materials datasets}

We apply the proposed architecture to three large-scale materials databases -- the MPtrj dataset~\citep{mptrj-and-chgnet}, the subsampled Alexandria dataset~\citep{alexandria,omat} and OMat24~\citep{omat} -- to demonstrate its behavior in the large-data regime and to compare it to state-of-the-art equivariant MLIPs. This ``PET-OAM'' model, trained on all three datasets in different stages (see App.~\ref{app:model}), contains 730M parameters. Thanks to the increased size of the node representations and the unconstrained architecture, this large parameter count can be achieved without sacrificing computational efficiency (see inference timings in Sec.~\ref{sec:spice}).

The model was pre-trained using non-conservative forces and stresses, then fine-tuned using their conservative counterparts. This two-step procedure was shown to save computational time in Refs.~\citep{darkside} and~\citep{esen}, often also resulting in better accuracies. 
Compared to other published training protocols~\citep{sevennet,grace,esen}, we find that our rotationally unconstrained architecture takes more passes through the data to train to convergence (although this does not necessarily translate to a higher computational cost, see Sec.~\ref{sec:spice}). 
This is not surprising: as argued in Sec.~\ref{sec:theory}, an unconstrained model needs to learn the equivariance constraints during training.
In contrast, trained unconstrained models can be fine-tuned quickly: for example, as shown in App.~\ref{app:fine-tuning}, a model trained on OMat24 can be fine-tuned to the smaller  Massive Atomistic Diversity (MAD) dataset~\cite{mazitov2025massiveatomicdiversitycompact} in fewer than 1/20th of the epochs needed to train from scratch~\cite{petmad}, while achieving 50\% lower test error than training on MAD alone.

\paragraph{Geometry optimization and phonon calculations from unconstrained models.}\label{ssec:geop-and-phonons}

\begin{figure*}[tb]
    \centering
    \includegraphics[width=1.0\linewidth]{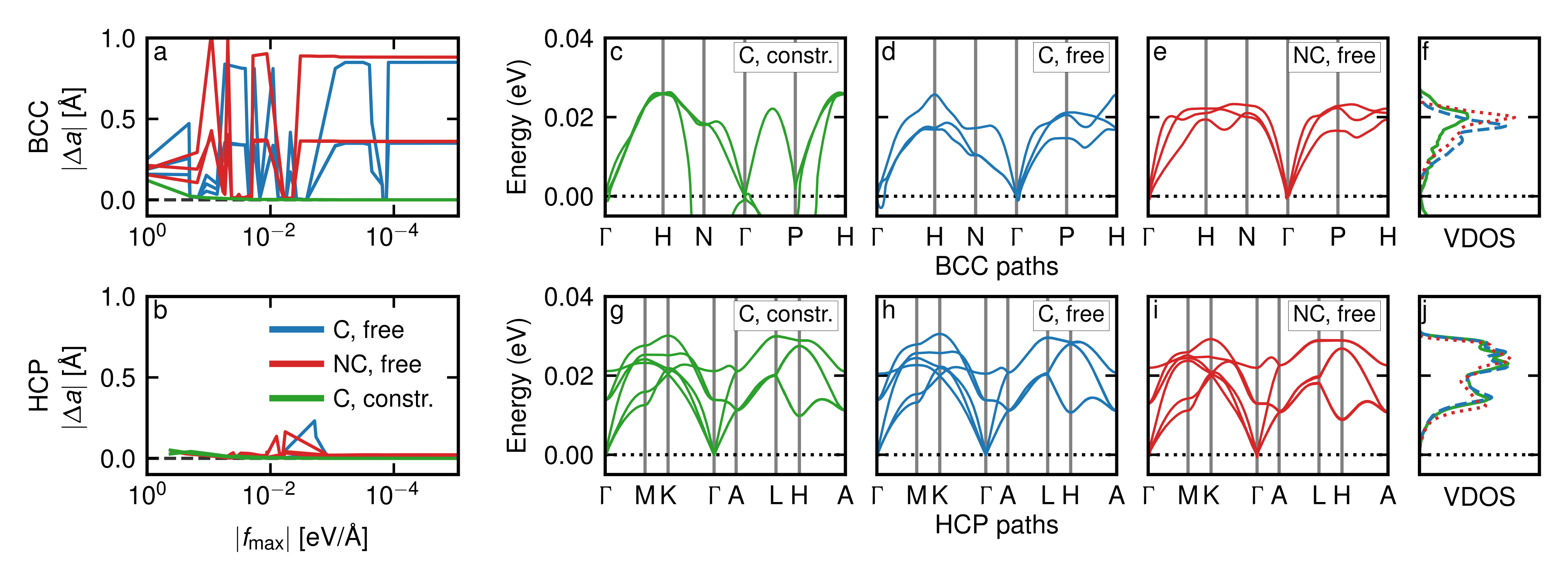}
    \vspace{-15pt}
    \caption{Geometry optimization and phonon calculations with PET-OAM. (a,b) Deviation of unit-cell lengths as a function of the maximum generalized force (force and stress component, as returned by ASE's ``FrechetCellFilter'') for BCC (a) and HCP (b) starting configurations. (c--e) Phonon bands of titanium starting from BCC, using conservative PET after constrained (c) and unconstrained (d) relaxation, and non-conservative PET after unconstrained relaxation (e). (g--i) Corresponding calculations starting from HCP. (f,j) Vibrational density of states for the two sets of calculations.}
    \label{fig:geomopt_and_phonons}
\end{figure*}

As a demonstrative example of the potential pitfalls of unconstrained models in static lattice calculations, we consider the case of elemental titanium. The stable phase of Ti at high temperature is body-centered cubic (BCC), but it is stabilized by entropic effects, and static calculations find the low-temperature hexagonal-closed-packed (HCP) to be preferred. 
In an equivariant model, symmetric but unstable structures such as BCC exhibit zero forces and scalar stress, so any relaxation initialized in BCC remains trapped within that symmetry.
Non-equivariant models are not bound by this constraint, and in fact residual asymmetries drive the relaxation toward more stable close-packed structures---in this case, close to the face-centered cubic (FCC) structure that can be obtained with a continuous deformation path.
Constraining the symmetry by projecting out the incompatible force components conserves the symmetry, for both conservative and non-conservative forces. 
Due to the presence of a small anisotropy in the stress, free relaxation leads to symmetry breaking---a small one for the stable HCP structure, and a dramatic deformation for the BCC cell, that relaxes to a locally stable geometry close to FCC (Figure~\ref{fig:geomopt_and_phonons}---note that residual instabilities in the fully relaxed phase come from unstable modes with wavelengths larger than the unit cell, and they vanish when using a supercell). 
Note that O(3)-averaging reduces but does not eliminate the anisotropy, and so it delays but does not prevent relaxation of the unstable BCC initial configuration.
We perform conservative and non-conservative relaxation for the 256\,963 structures within the WBM dataset~\citep{wbm}, with a convergence threshold of 0.05\,eV/\AA. 
We find that 70\%{} (16\%{}) of the final configurations are detected to retain the symmetry of the the initial structure with the default threshold of 0.01\,\AA~for the conservative (non-conservative) model, and 88\%{} (81\%{}) with a threshold of 0.1\,\AA. 
Notably, O(3)-averaging increases the number of symmetric structures to 77\%{} with symprec 0.01\,\AA{} in the non-conservative case.
We therefore recommend to use looser convergence tolerances or to enforce rotational averaging whenever performing free geometry optimizations with unconstrained models, especially when using direct-force predictions.

The outcomes of a phonon calculation depend mostly on the relaxed geometry. Using a constrained BCC structure leads (as it should) to unstable phonon modes, whereas the relaxed FCC structure is a local minimum and displays no unstable lattice vibrations. 
When considering the stable HCP configuration, instead, the phonon dispersion curves are almost identical regardless of whether or not the structure has been free to relax to a slightly symmetry-broken geometry. We reiterate that automatic workflows might choose a dramatically different BZ sampling path if a lower symmetry is detected. This is inconvenient, but has no consequence on physical properties such as the total vibrational density of states, that depend on an integral over the entire BZ.
Even when symmetrizing the Hessian and removing the total force component, non-conservative models tend to yield significantly different phonon dispersion curves. Increasing the finite-difference displacements used to estimate the Hessian helps stabilize calculations, but at present we consider it to be safer to use a conservative model when evaluating phonons.
These demonstrative examples are representative of the general behavior of unconstrained models for lattice dynamics calculations. An accurate conservative, rotationally unconstrained model can be used safely---as will be clear from the excellent performance on benchmarks that use vibrational properties---although care must be taken to avoid qualitative changes in the detected symmetry.

\paragraph{The matbench-discovery benchmark}

\begin{table}[t]
\caption{Model performance on selected metrics from the matbench-discovery benchmark suite. We include the top 5 openly available models from the benchmark leaderboard at the time of writing. For each metric, the best model is highlighted in bold and the second best is underlined.}
\label{tab:matbench}
\begin{center}
\begin{tabular}{l|ccccc}
\toprule
Model & DAF($\uparrow$) & Acc.($\uparrow$) & F1($\uparrow$) & $\kappa_\textrm{SRME}$($\downarrow$) & RMSD($\downarrow$)\\
\midrule
PET                   & \B{6.075} & \B{0.977} & \U{0.924} & \B{0.119} & \B{0.060} \\
\midrule
eSEN~\cite{esen}                  & \U{6.069} & \B{0.977} & \B{0.925} & 0.170 & 0.061 \\
NequIP~\cite{nequip}              & 5.869 & 0.971 & 0.906 & \U{0.125} & 0.063 \\
MatRIS~\cite{matris}              & 6.039 & 0.976 & 0.921 & 0.218 & \B{0.060} \\
SevenNet~\cite{sevennet}          & 5.954 & 0.971 & 0.906 & 0.192 & 0.062 \\
\bottomrule
\end{tabular}
\end{center}
\end{table}

\begin{table}[b]
\caption{Test set accuracies for models trained on the dataset presented in Ref.~\citep{mace-off}. Energy MAEs are shown in units of meV per atom; force MAEs are shown in units of meV/\AA. For each metric, the best model is highlighted in bold and the second best is underlined.}
\label{tab:spice}
\begin{center}
\begin{tabular}{l|cc|cc|cc|cc}
\toprule
\multicolumn{1}{c}{Subset} & \multicolumn{2}{c}{MACE} & \multicolumn{2}{c}{EScAIP} & \multicolumn{2}{c}{eSEN} & \multicolumn{2}{c}{PET} \\
\midrule
 & E & F & E & F & E & F & E & F \\
\midrule
PubChem               & 0.88 & 14.75 & 0.53 &  5.86 & \U{0.15} & \U{4.21} & \B{0.09} & \B{3.53} \\
DES370K Mon.      & 0.59 &  6.58 & 0.41 &  3.48 & \U{0.13} & \U{1.24} & \B{0.10} & \B{1.00} \\
DES370K Dim.        & 0.54 &  6.62 & 0.38 &  2.18 & \U{0.15} & \U{2.12} & \B{0.12} & \B{1.20} \\
Dipeptides            & 0.42 & 10.19 & 0.31 &  5.21 & \U{0.07} & \U{2.00} & \B{0.05} & \B{1.55} \\
Solv. Aminoacids  & 0.98 & 19.43 & 0.61 & 11.52 & \U{0.25} & \B{3.68} & \B{0.17} & \U{4.37} \\
Water                 & 0.83 & 13.57 & 0.72 & 10.31 & \U{0.15} & \B{2.50} & \B{0.13} & \U{3.05} \\
QMugs                 & 0.45 & 16.93 & 0.41 &  8.74 & \U{0.12} & \U{3.78} & \B{0.08} & \B{2.91} \\
\bottomrule
\end{tabular}
\end{center}
\end{table}

\begin{figure*}[t]
    \centering
    \includegraphics[width=\linewidth]{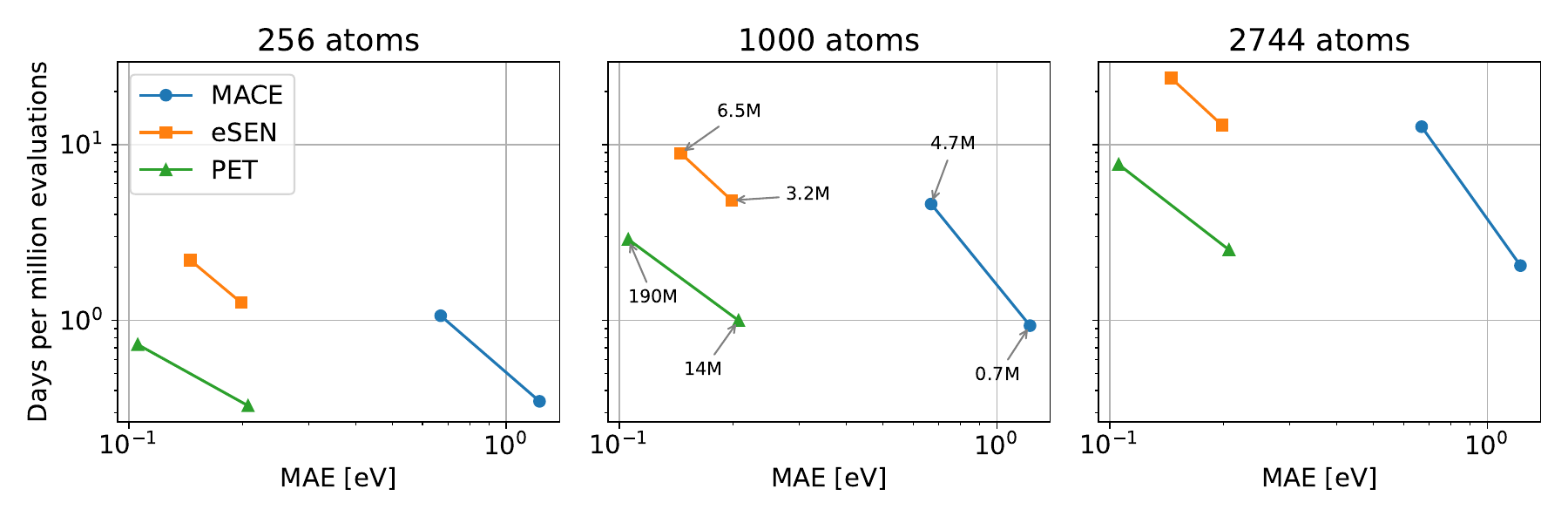}
    \vspace{-22pt} %
    \caption{Accuracy-speed Pareto front for models trained on the SPICE dataset, at varying structure sizes. The energy error on the x-axis is the average of all subset-specific errors. Model sizes for the six benchmarked models are shown in the middle panel. More details are available in App.~\ref{app:spice}.}
    \label{fig:spice}
\end{figure*}

Although public benchmarks should not be taken as the sole metric to assess the usefulness of a model, they provide an objective scale on which one can compare different architectures. 
Among material-based benchmarks for MLIPs, matbench-discovery~\citep{matbench} features the widest variety of architectures. This benchmark emphasizes metrics that reflect the ability of a model to be used for materials discovery, and especially its ability to detect the most stable configurations among those of similar density or composition through a convex hull analysis. 
We focus on the following metrics, which are useful predictors of performance in material discovery workflows: (1) DAF (discovery acceleration factor): the acceleration factor of material discovery through the model compared to brute-force search; (2) accuracy: this is the fractional accuracy of the exercise if seen as a binary classification task (stable/unstable); (3) F1 score: harmonic mean of precision and recall for the binary classification task. 
Besides these three metrics, we will also use (4) an RMSD metric for geometry optimization (root-mean-square deviation of the MLIP-optimized structure compared to DFT) and (5) $\kappa_\textrm{SRME}$, which quantifies the quality of phonons calculated from the model \citep{kappa_srme}.
The results (\autoref{tab:matbench}) show that the methods proposed in the current work reach state-of-the-art accuracy on materials discovery workflows, as well as on the calculation of static properties at local minima (phonons and optimized geometries). All results we report here are obtained after rotational symmetrization by averaging over an SO(3) grid, although this has a significant effect only on the RMSD values. More details on the matbench-discovery evaluation are available in App.~\ref{app:matbench}, where we also show results without rotational symmetrization and/or using non-conservative forces. In particular, the non-conservative heads of our models perform surprisingly well compared to previous non-conservative results in the public domain~\cite{matbench}. Further benchmarking on LAMbench~\citep{lambench} and MADBench~\citep{petmad} is available in App.~\ref{app:lambench}~and~\ref{app:madbench}, confirming the very strong performance of the models proposed in this work on materials.

\subsection{The SPICE molecular benchmark}\label{sec:spice}

Unlike the material domain, applications of MLIPs to molecular systems have not yet seen the development of established benchmark suites. Two of the most widely used large-scale MLIP benchmarks in the molecular domain are the test accuracy on the SPICE~\citep{spice} dataset (in the version proposed by Ref.~\citep{mace-off}) and the Open Molecules leaderboard and community challenges~\cite{omol}. 
In this section, we focus on the former; training and evaluation exercises on the large-scale OMol-1 dataset~\cite{omol} are available in App.~\ref{app:omol}.

\paragraph{Training and accuracy}

We train a 190M-parameter model on SPICE, using conservative forces throughout and training for three times the number of epochs reported by eSEN~\citep{esen}. Although training times for the latter are not available, the inference timings in Fig.~\ref{fig:spice} would suggest that the two models required around the same amount of compute to train. The proposed unconstrained architecture exceeds the accuracy of state-of-the-art models on this benchmark (Table~\ref{tab:spice}), demonstrating its remarkable accuracy in the molecular domain.

\paragraph{Inference timings}

Molecular force fields are often used to perform molecular dynamics simulations, making inference efficiency of primary importance. 
Unlike high-throughput workflows, molecular dynamics is not trivially parallelizable. 
Furthermore, large-scale simulations are often needed in this domain to fully capture the structural complexity of biochemical processes, increasing the computational cost of inference. Fig.~\ref{fig:spice} shows three accuracy-speed Pareto fronts for the models presented in Table~\ref{tab:spice} for systems of varying numbers of atoms. From these experiments, it can be seen that the unconstrained models presented in this work constitute an excellent compromise between accuracy and inference speed. More details on this benchmark are available in App.~\ref{app:spice}.

\section{Discussion}

In this work, we have shown that unconstrained models (in the rotational sense and/or in the direct-force sense) can be scaled successfully to train on large datasets and afford accuracies and speed on par with state-of-the-art equivariant architectures for MLIPs.
From our investigation it appears that, when compared to physically constrained models, non-equivariant models must train for a larger number of epochs to converge training, although the lower cost largely compensates for that.
Direct force models, on the other hand, show accelerated training (as also observed in Refs.~\citep{esen} and~\citep{darkside}) and can be fine-tuned to obtain conservative models at a much reduced computational effort.

At inference time, we have demonstrated that rotationally unconstrained models can be more efficient than equivariant models, although further tuning of either class of architectures, as well as software optimization, might shift the balance in both directions.
Non-conservative models consistently show the expected 2-3$\times$ theoretical speed-up over their conservative counterparts.
Even though we were able to establish that rotationally unconstrained models at scale can match the predictive accuracy of equivariant models on downstream tasks (at times with minor modifications to the evaluation procedure), non-conservative force models with equal or better test set accuracies compared to conservative models often fail to achieve the same quantitative performance. We conclude that they are best used as a pre-training step in the training of conservative models~\citep{darkside,esen,uma} or in combination with conservative models for finite-temperature applications~\citep{darkside}.

Overall, this work shows that the PET architecture, which had already proven itself as a lightweight model trained on a comparatively small dataset~\cite{petmad}, retains excellent performance at scale.
It establishes rotationally unconstrained architectures as promising alternatives to achieve better trade-offs between accuracy, parameter budget, and computational cost in general-purpose MLIPs and their applications. 
The pre-trained models we make available with this work span a broad range of parameter counts and computational cost, and can be used by practitioners to quickly fine-tune models on bespoke datasets, facilitating applications, as well as testing the PET architecture on more challenging use cases beyond the benchmarks we present.  
Given that, in this domain, unconstrained models have received much less attention than equivariant neural networks, we are convinced that there is substantial potential for further improvements in both accuracy and efficiency.

\section{Data availability statement} 

The data that support the findings of this study are openly available. The PET models are freely available at \url{https://github.com/lab-cosmo/upet/}, and they can be run and fine-tuned using tools available from \url{https://metatensor.org/}, which are also documented in Ref.~\citenum{metatensor}.
Examples of how to run geometry optimization and phonon calculations with unconstrained models can be found at \url{https://atomistic-cookbook.org/examples/pet-relaxation/pet-relaxation.html} and \url{https://atomistic-cookbook.org/examples/pet-phonons/pet-phonons.html}, respectively.

\begin{acknowledgments}
The authors would like to thank M. Langer, B. Wood and M. Gao for helpful discussions.

AM and MC acknowledge support from an Industrial Grant from BASF and from EPFL.
FB was supported by a project within the Platform for Advanced Scientific Computing (PASC), and by the MARVEL National Centre of Competence in Research (NCCR), funded by the Swiss National Science Foundation (SNSF, grant number 182892)
PP and MC acknowledge the funding from the European Research Council (ERC) under the European Union’s Horizon 2020 research and innovation programme (grant agreement No 101001890-FIAMMA). This work was supported by the Swiss AI Initiative (2025 Fellowship Program). JS acknowledges support from the Swiss National Science Foundation (SNSF) under Ambizione grant number 233444. Computing resources were also provided by the Swiss National Supercomputing Centre under project no. lp26, lp95 and lp133.
\end{acknowledgments}

\bibliographystyle{unsrt}

\clearpage

\renewcommand{\thefigure}{A\arabic{figure}}
\setcounter{figure}{0}
\renewcommand{\thesection}{A\arabic{section}}
\setcounter{section}{0}

\onecolumngrid
\onecolumngrid

\centerline{\huge \bfseries Appendices}
\appendix
\section{Model and training details}\label{app:model}

\subsection{Architecture}

The proposed architecture is based on a modified version of the original PET from Ref.~\citep{pozd-ceri23nips}, and it is implemented in the metatrain library~\citep{metatensor}.
Besides the changes to the architecture described in the main text, a number of further smaller modifications were made, namely:
\begin{itemize}
    \item the use of skip connections instead of summing layer-wise predictions
    \item the use of more processed representations to make predictions
    \item a multi-layer perceptron is used to combine $ij$ and $ji$ representations during message passing, as opposed to a plain sum
\end{itemize}
These, along with the changes highlighted in the main text, are illustrated in Fig.~\ref{fig:pet}. All proposed modifications improve the accuracy of training on diverse datasets. As an example, we report a 10-15\% improvement in force MAE on the MAD dataset in Fig.~\ref{fig:finetune} compared to literature results, with comparable computational efficiency. However, we do not exclude that further tuning of the architecture might improve accuracies and/or inference timings further.

\resub{Finally, we would like to stress that all experiments presented in this manuscript do not make use of the Equivariant Coordinate System Ensemble (ECSE)~\cite{pozd-ceri23nips} to enforce exact energy invariance.}

\subsection{Adaptive cutoff}\label{app:adaptive-cutoff}

\begin{figure}[b]
    \centering
    \includegraphics[width=0.7\linewidth]{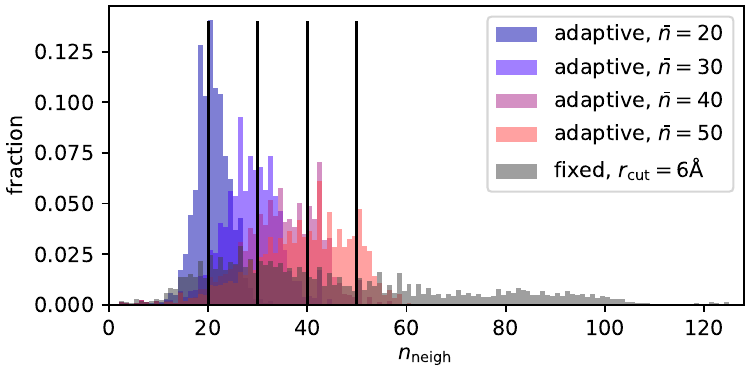}
    \caption{Histogram of the number of neighbors selected by the adaptive cutoff algorithm with different target $\bar{n}$, and compared with that for a fixed cutoff of 6\AA{}. 
    The dataset is a collection of 2700 structures randomly selected from the highly diverse MAD dataset~\citep{mad}.}
    \label{fig:adaptive}
\end{figure}

Diverse datasets that span the full periodic table include materials with wildly varying densities. This is problematic because the number of edges included within a fixed cutoff varies wildly. To ensure that the computational and parameter budget is allocated more uniformly across different structures, and that data can be packed more tightly in dense storage structures, it is useful to let the cutoff vary to include a roughly-constant number of neighbors. 
To achieve an adaptive cutoff in an efficient and simple manner, and such that the cutoff value depends smoothly on the coordinate of the atoms, we re-use some of the ideas introduced in Ref.~\citenum{pozd-ceri23nips}. 
We define a large cutoff value $r_\text{max}$, over which all interatomic distances $r_{ij}$ are computed. For each atom $i$ in a structure, we then compute the number of neighbors within a grid of probe cutoff distances, 
\begin{equation}
n_k(A_i) = \sum_j f_\text{c}(r_ij, r_k, \Delta r),
\end{equation}
where we use the smooth, infinitely differentiable function
\begin{equation}
f_c(r, r_\text{c}, \Delta r) = \left\{
\begin{array}{ll}
   0  &  r\ge r_\text{c} \\
   1  &  r\le r_\text{c} -\Delta r \\
   \frac{1}{2}\left[1+\tanh \frac{1}{\tan[\pi (r+\Delta r-r_\text{c})/\Delta r]}  \right] &
   \text{elsewhere}
\end{array}
\right.    
\end{equation}
We then look for the cutoff value that would encompass a target number $\bar{n}$ of neighbors, using the expression
\begin{equation}
\bar{r}(A_i) = \sum_k r_k g((n_k-\bar{n}+B(r_k))/\Delta n) /\sum_k g((n_k-\bar{n}+B(r_k))/\Delta n).
\end{equation}
$B(r)=\bar{n}(r/r_\text{max})^3$ is a ``baseline'', corresponding to a uniform atom density integrating to $\bar{n}$ at the maximum cutoff, that we add to the estimated number of neighbors and serves multiple purposes. It stabilizes the evaluation of $\bar{r}$ in cases where there are fewer atoms within the outer cutoff than the target number and reduces the adaptive cutoff value in cases where there are many atoms close to the outer cutoff, avoiding selecting a number of actual active edges that is much larger than the target. 
The probe cutoff radii can be computed over a coarse grid; we use a spacing $\delta r=\Delta r/4$ as a heuristic.
The Gaussian width used to determine the adaptive cutoff value should be large enough to interpolate smoothly between grid points. Since the optimal width depends on the value of the estimated neighbor counts in the neighboring probe radii, we find that an effective strategy is to use an adaptive value $\Delta n_k = (n_{k+1}+B(r_{k+1})-n_{k-1}-B(r_{k-1}))/2$, computed as centered differences with respect to the neighbor counts (including the baseline correction, and using one-sided expressions at the edges). Finally, the weighting of a given edge is given by the mean of the weights of the corresponding two nodes to ensure symmetry, and edges with a weight of zero are discarded. These weights are then reused as the weights in PET's attention mechanism (see Ref.~\cite{pozd-ceri23nips}).

\subsection{Training}

All training and fine-tuning runs were executed with a standard cosine learning rate scheduler, with a linear warm-up stage corresponding to 10\% of the total number of training steps. All training runs were executed in single-precision, without using TensorFloat-32 operations. A chemical-composition-based linear model was fitted for the energies and removed from the energy targets before the fitting exercises, except for fine-tuning runs. Furthermore, scaling of all targets to unit standard deviation across the training set was employed.
A standard Huber loss was used for all training exercises, except for fine-tuning on the MAD dataset (Sec.~\ref{app:fine-tuning}), where a MSE loss was used. The total loss is composed of a sum of individual terms for per-atom energies, forces and per-atom virials (we use stresses instead in the non-conservative case). Rotational and inversion-based data augmentation was employed for all training runs, according to the symmetry of the different targets (energy, forces, stress/virial). Gradient norm clipping to a value of 1.0 was used throughout. It should be noted that all conservative fine-tuning runs keep training the non-conservative force and stress outputs with a small weight in the loss function. This is done to allow multiple-time-stepping simulations using both conservative and non-conservative heads \citep{darkside}.

For the fine-tuning of the ``OAM'' model, we rejected all batches with more than 350 atoms (in a distributed setting, this implies rejection of all batches across all processes if even only one local batch is too large). This constraint, together with the adaptive cutoff strategy detailed in App.~\ref{app:adaptive-cutoff}, implicitly sets a limit to the memory usage per batch during training. In general, this allows one to push the size and parameter counts of the model on heterogeneous datasets without recurring to alternative strategies such as model sharding across devices.

The official training/validation/test splits were used for the OMat24~\citep{omat}, sAlex~\citep{omat}, SPICE~\citep{mace-off}, MAD~\citep{mad} and OMol25~\cite{omol} datasets. Since the MPtrj dataset~\citep{mptrj-and-chgnet} does not have official splits, as far as we are aware, we added it in its entirety to the training set for the OAM fine-tuning run.

\resub{See Sec.~\ref{app:omol} for model and training details relative to the OMol models.}

\clearpage\subsection{Hyperparameters}\label{app:hypers}

The hyperparameters employed to train the models presented in this work are shown in Tab.~\ref{tab:hyperparameters}. \resub{See Sec.~\ref{app:omol} for model and training hyperparameters relative to the OMol models.}

\begin{table}[ht]
    \centering
    \caption{Model and training hyperparameters for the models evaluated in this work. A linear learning-rate warm-up of 10\% of the total number of epochs was performed in all cases. The number of node features corresponds to four times that of edge features for all models. See Sec.~\ref{app:omol} for model and training hyperparameters relative to the OMol models.}
    \label{tab:hyperparameters}
    \begin{tabular}{l|cc|c|cc}
    Model & OMat-NC & OMat-C & OAM & SPICE-L & SPICE-S\\
    \midrule
    Training starts from & scratch & OMat-NC & OMat-C & scratch & scratch \\
    Parameter count & 730M & 730M & 730M & 190M & 14M \\
    Conservative training & no & yes & yes & yes & yes \\
    Number of epochs & 10 & 5 & 1 & 300 & 300 \\
    Edge features & 640 & 640 & 640 & 512 & 192 \\
    GNN layers & 5 & 5 & 5 & 3 & 3 \\
    Attention layers & 3 & 3 & 3 & 2 & 1 \\
    Graph cutoff radius (\AA) & 10.0 & 10.0 & 10.0 & 4.5 & 4.5 \\
    Adaptive neighbor number & 40 & 40 & 40 & -- & -- \\
    Max learning rate & 2e-4 & 1e-4 & 5e-6 & 2e-4 & 2e-4 \\
    Batch size & 2048 & 2048 & 512 & 128 & 128 \\
    Max atoms per batch & -- & -- & 350 & -- & -- \\
    E loss weight &       0.01 & 1.0 & 1.0 & 1.0 & 1.0 \\
    F loss weight &       1.0  & 1.0 & 1.0 & 1.0 & 1.0 \\
    S/V loss weight &     0.01 & 1.0 & 1.0 & -- & --\\
    E Huber threshold &   0.015 & 0.015 & 0.010 & 0.003 & 0.003 \\
    F Huber threshold &   0.010 & 0.040 & 0.050 & 0.004 & 0.004 \\
    S/V Huber threshold & 0.005 & 0.030 & 0.050 & -- & -- \\
    \end{tabular}
\end{table}

\clearpage
\subsection{Pareto frontier for PET-OMat models}

\begin{table}[ht]\centering 
\label{tab:omat-hyperparameters}
 \caption{Model hyperparameters for the PET-OMat models. The ``OMat-XL'' model we present here is what we call "OMat" elsewhere.}
    \begin{tabular}{l|ccccc}
    Model & OMat-XS & OMat-S & OMat-M & OMat-L & OMat-XL \\
    \midrule
    Parameter count & 4.5M & 25.9M & 109M & 255M & 730M \\
    Edge features & 128 & 256 & 384 & 512 & 640 \\
    GNN layers & 2 & 3 & 3 & 4 & 5 \\
    Attention layers & 1 & 1 & 2 & 2 & 3 \\
    Graph cutoff radius (\AA) & 7.5 & 8.0 & 8.5 & 9.0 & 10.0 \\
    Adaptive neighbor number & 8 & 16 & 24 & 32 & 40 \\
    \end{tabular}
\end{table}

In addition to the very large model that we use to demonstrate the ability of the PET architecture to scale to large parameter counts, we perform a series of training exercises on the OMat24 dataset~\citep{omat} using different sets of hyperparameters (Table~\ref{tab:omat-hyperparameters}).
Besides their value in the context of an ablation study on model complexity, we envisage these models to be used as foundation models to be fine-tuned on more focused and/or curated datasets, and we chose hyperparameters so as to span a broad range of cost/accuracy tradeoff scenarios.

\begin{figure}[ht]
    \centering
    \includegraphics[width=0.6\linewidth]{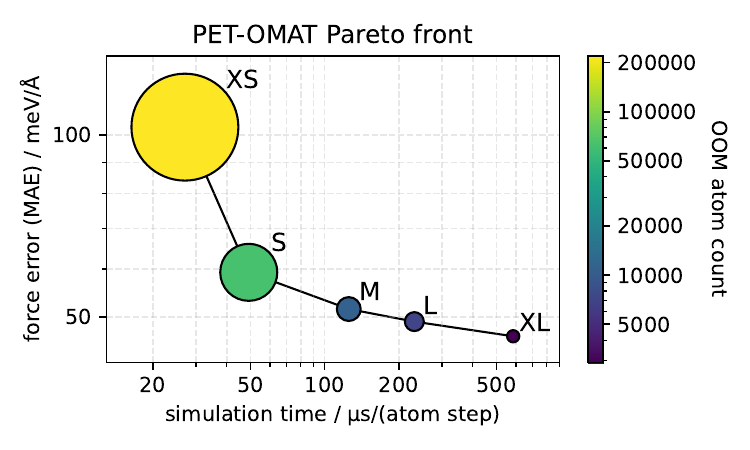}
    \caption{Pareto plot of accuracy (force MAE on a hold-out test set) and computational cost (measured from ASE~\citep{ase-paper}) for a series of PET models trained on the OMat24 dataset. The maximum system size that can be reached before hitting an out-of-memory error on the hardware we used (a single NVIDIA H100 GPU with 92 GB of memory) is visualized as the color and size of the data points.
    }
    \label{fig:omat-pareto}
\end{figure}

\section{Fine-tuning on smaller datasets}\label{app:fine-tuning}

\begin{figure}[ht]
    \centering
    \includegraphics[width=\linewidth]{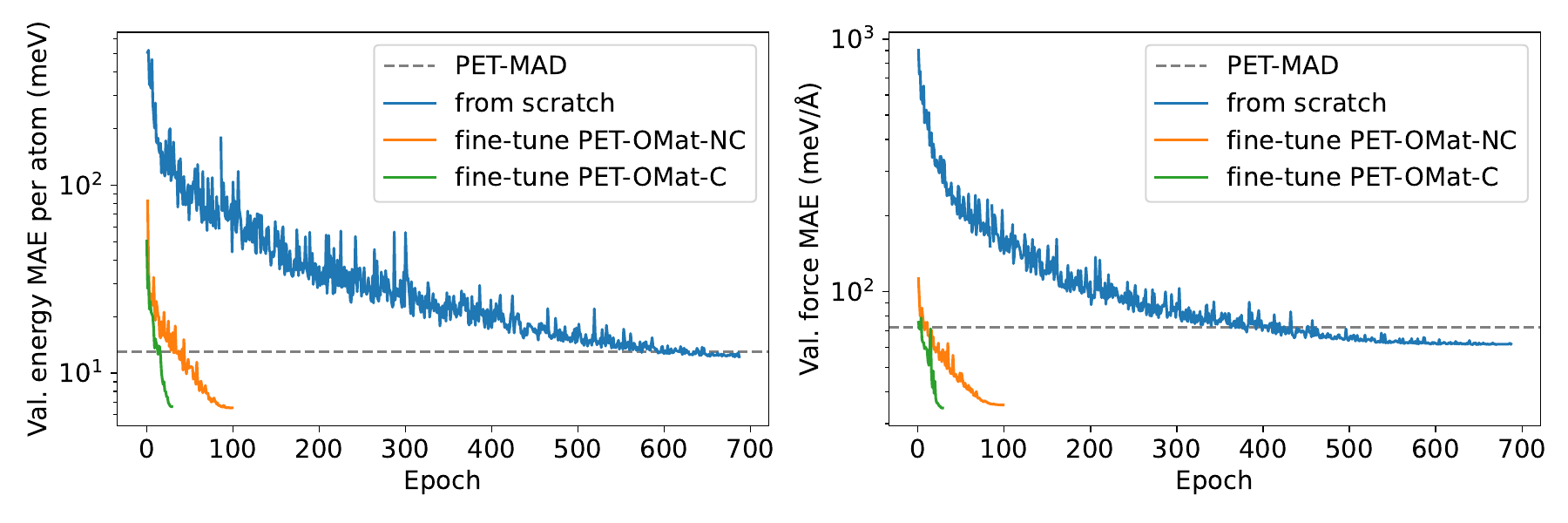}
    \caption{Validation set accuracy of the PET-MAD~\citep{petmad} model, a model trained from scratch on the MAD dataset using the architecture proposed in this work, and fine-tuning from a model trained on the OMat24 dataset.}
    \label{fig:finetune}
\end{figure}

As a demonstration of the fine-tuning capabilities of unconstrained architectures, we consider fine-tuning of two early versions of the models we trained on the OMat24~\citep{omat} dataset: PET-OMat-NC (non-conservative) and PET-OMat-C (conservative). We fine-tune these models on the MAD~\citep{mad} dataset, which is smaller by more than 1000 times, but which was nonetheless used to train a successful general-purpose interatomic potential in Ref.~\citep{petmad}. We compare the fine-tuned model with training from randomly initialized weights on the same dataset and with the reported accuracy from Ref.~\citep{petmad} in Fig.~\ref{fig:finetune}.

Fine-tuning our OMat models leads to halving of the validation energy and force errors compared to training from scratch. Furthermore, training time is greatly reduced: we found that it is possible to achieve near-converged validation metrics by fine-tuning for 100 epochs on a non-conservative OMat model and 30 epochs on a conservative OMat model, as opposed to 700 or more when training from scratch. We postulate that, once equivariance is learned on a larger datasets, fine-tuning is efficient and largely unaffected by any potential training-time slowdowns due to learning equivariance. This demonstrates the benefits of using our pre-trained models to achieve better accuracies on smaller datasets, whose targets can be computed consistently at high levels of theory (i.e., using more expensive approximations of quantum mechanics).
In Sec.~\ref{app:madbench}, we show that this ``OMAD'' model, when evaluated consistently, shows superior accuracy and transferability compared to OAM models.

\section{Matbench-discovery evaluation}\label{app:matbench}

Evaluation on matbench-discovery~\citep{matbench} is executed through scripts available in its public GitHub repository (https://github.com/janosh/matbench-discovery). At the time of writing, these differ only marginally from standard scripts used to evaluate other architectures.

\begin{table}[h]
\caption{Performance comparison of conservative and non-conservative models on the matbench-discovery benchmark, with and without rotational averaging.}
\label{tab:c-vs-nc}
\begin{center}
\begin{tabular}{l|ccccc}
\toprule
Model & DAF($\uparrow$) & Acc.($\uparrow$) & F1($\uparrow$) & $\kappa_\textrm{SRME}$($\downarrow$) & RMSD($\downarrow$)\\
\midrule
PET-OAM (C, avg)  \,            & 6.075 & 0.977 & 0.924 & 0.119 & 0.060 \\
PET-OAM (C, raw)  \,            & 6.018 & 0.975 & 0.921 & 0.119 & 0.070 \\
\midrule
PET-OAM (NC, avg) \,            & 6.100 & 0.976 & 0.922 & 0.197 & 0.072 \\
PET-OAM (NC, raw) \,            & 5.968 & 0.973 & 0.913 & 0.216 & 0.099 \\
\bottomrule
\end{tabular}
\end{center}
\end{table}

\paragraph{Effect of conservative predictions and rotational averaging}
Optionally, we employ SO(3)-averaging over a Lebedev grid with $L=3$ and/or the non-conservative force and stress heads of our models. We report the results in Tab.~\ref{tab:c-vs-nc}. While we do not find symmetrization to have an impact on the phonon-related task in the benchmark suite (at least in the conservative case), the effects are small but noticeable on materials discovery tasks, and sizable on geometry optimization accuracies. Although we were able to obtain relatively good metrics from non-conservative models (especially on $\kappa_\textrm{SRME}$ when compared to the literature, see also the next paragraph), we find that non-conservative models consistently perform worse on the downstream tasks considered in matbench-discovery, despite them often achieving better training and validation accuracies if trained from scratch.

\paragraph{Effect of the finite-difference displacement on phonon calculations}

Tab.~\ref{tab:finite-displacement} reports the performance of different models on the phonons task of matbench-discovery, which evaluates the symmetric mean relative error on lattice thermal conductivity~\citep{pota2025thermal}, for varying finite-difference displacements in force-constant calculations. All results use central finite differences. We find that this choice is crucial for non-equivariant models (and it is also adopted for ORB models), as it can more than halve the error by averaging out small inconsistencies between symmetry-equivalent displacements (see also App.~\ref{sec:phonons}). 
Our results show little variability with the displacement size, with optimal values slightly larger than the default 0.03\,\AA{} used in phonopy (and matbench-discovery). A similar investigation in Ref.~\citep{esen} also found that the 0.03 \AA~standard is not optimal in general. Overall, the conservative model achieves state-of-the-art performance, while the non-conservative model remains competitive with conservative approaches and outperforms other direct-force models on the leaderboard.

\begin{table}[ht]
\centering
\caption{Effect of finite-different displacement size on the $\kappa_{\mathrm{SRME}}$ metrics of matbench-discovery.}
\label{tab:finite-displacement}
\begin{tabular}{lccccc}
\toprule
\multirow{2}{*}{Model} & \multicolumn{5}{c}{Finite-difference displacement (\AA)}\\
 & 0.01 & 0.03 & 0.05 & 0.07 & 0.10 \\
\midrule
PET-OAM (C, raw) & 0.157 & 0.119 & \textbf{0.110} & \textbf{0.110} & 0.124 \\
PET-OAM (NC, raw) & \textbf{0.216} & \textbf{0.216} & 0.218 & 0.226 & 0.253 \\
\bottomrule
\end{tabular}
\end{table}

\section{LAMbench evaluation}\label{app:lambench}

Although matbench-discovery~\citep{matbench} remains the most popular public benchmark for atomistic models in the materials domain, it is important to consider other available benchmarks to obtain a broader perspective of the performance of various models on a more diverse set of tasks. We evaluated the performance of our models using the recent LAMBench benchmark \citep{lambench}, which focuses on probing the generalizability, inference speed, and stability of universal models in atomistic simulations. For this benchmark, in addition to PET-MAD~\citep{petmad} and the PET-OMAD model described in App.~\ref{app:fine-tuning}, we used a set of preliminary conservative and non-conservative models that we trained on the OMat24, subsampled Alexandria, and MPtrj datasets. When interpreting the results below, \textbf{it is important to keep in mind that there is usually no consistency between the DFT settings used to prepare the benchmark data and model training sets}. This lack of consistency can severely affect the quality of predictions because it introduces unavoidable and unlearnable noise in the target data, which cannot be separated from the real approximation errors of the models (more detail on this effect can be found in Ref.~\citep{petmad}). Therefore, the results in this section should only be used for qualitative model assessment. Generally, we recommend benchmarks which use consistent levels of DFT in order to perform quantitative assessments (see Sec. \ref{app:madbench} below). 
We first run the force-field generalizability test resolved by three separate categories: Molecules, Inorganic Materials and Catalysis. In each category, the accuracy of each model in predicting energies, forces and stresses (if available) is evaluated across several out-of-domain category-specific datasets, and then the overall performance score is computed based on the averaged normalized error values. All technical details on the score calculation procedure can be found in Ref.~\citep{lambench}.

The results of the evaluation are shown in Table \ref{tab:lambench}. Among all models trained in this work, PET-OAM-C demonstrates the best score across all three domains, showing the best overall results on Molecules, and reaching the same accuracy as the DPA-3.1-3M on Inorganic Materials. In the Catalysis domain, both conservative and non-conservative PET-OAM models show second-best result, outperforming all other OAM-trained models in the list. The superior performance of the DPA-3.1-3M model in this case most likely originates from the presence of explicit catalytic data in the training set~\citep{DPA}. PET-OMAD performs notably worse compared to PET-OAM, and even falls behind the PET-MPtrj on Inorganic Materials, despite being pre-trained on the large OMat24 dataset. This effect stems from differences in ab initio theory levels used in the MAD dataset (on which PET-OMAD is ultimately fine-tuned) and other datasets, which usually have settings consistent (or nearly consistent) to those in the MPtrj dataset~\citep{mptrj-and-chgnet} or the Alexandria dataset~\citep{alexandria}, and therefore suffer much less from the difference in the theory baseline. As we demonstrate in Sec.~\ref{app:madbench}, PET-OMAD outperforms other models upon consistent evaluation. A more detailed discussion of the effect of the consistency of DFT settings in model evaluation can be found in Ref.~\citep{petmad}. 

\begin{table}[t]
\caption{
Evaluation of various universal MLIPs on the force-field generalizability test of the LAMBench benchmark from Ref.~\citep{lambench}. Performances of PET-OAM, PET-MPtrj and PET-OMAD are computed in this work, and other results are reproduced from the reference paper~\citep{lambench}. The domain-specific score ranging from 0 to 1 is computed based on model errors in predicting energies and forces. For each metric, the best model is highlighted in bold and the second best is underlined.
}
\label{tab:lambench}
\begin{center}

\begin{tabular}{l|ccc}
\toprule
\multirow{2}{*}{Model} & \multicolumn{3}{c}{Task} \\
& Molecules & Inorganic Materials & Catalysis \\
\midrule
PET-OAM-C     & \B{0.85} & \B{0.84} & \U{0.74} \\
\midrule
PET-OAM-NC    & 0.78 & 0.79 & \U{0.74} \\
PET-MPtrj-C         & 0.80 & \U{0.83} & 0.48 \\
PET-MPtrj-NC        & 0.69 & 0.77 & 0.46 \\
PET-MAD          & 0.79 & 0.76 & 0.44 \\
PET-OMAD      & 0.81 & 0.78 & 0.58 \\
\midrule
DPA-2.4-7M       & 0.82 & 0.80 & 0.66 \\
DPA-3.1-3M       & \U{0.84} & \B{0.84} & \B{0.79} \\
MACE-MP-0        & 0.64 & 0.72 & 0.59 \\
MACE-MPA-0       & 0.66 & 0.79 & 0.62 \\
Orb-v2           & 0.78 & 0.75 & 0.70 \\
Orb-v3           & 0.82 & 0.82 & 0.71 \\
SevenNet-l3i5    & 0.75 & 0.76 & 0.51 \\
SevenNet-MF-ompa & 0.75 & \U{0.83} & 0.66 \\
MatterSim-v1-5M  & 0.74 & 0.82 & 0.59 \\
GRACE-2L-OAM     & 0.75 & 0.82 & 0.67 \\

\bottomrule
\end{tabular}

\end{center}
\end{table}

In addition to force-field generalizability, we also perform the property prediction test, which includes reaction and activation energy prediction in catalytic reactions (OC20-NEB task), elastic constant prediction (elastic task), and conformer energy prediction for a selected set of organic molecules (Wiggle150 task). The results are shown in Tables \ref{tab:lambench-neb}, \ref{tab:lambench-elastic}, and \ref{tab:lambench-wiggle150}, respectively. The lack of explicit molecular and surface data in the OAM dataset lead to a dramatic decrease in PET-MP and PET-OAM model accuracy compared to bulk systems. At the same time, almost all PET models show high (and sometimes the best) success rates, which means that, in most cases, their MAE in predicting $E_a$ is less than 0.1 eV. This combination is most likely caused by outliers in the data, which can corrupt the overall MAE values, while preserving the fraction of successful runs on a decent level.

At the same time, PET-OMAD achieves the third-best results in predicting reaction energies and barriers, closely competing with DPA-2.4-7M, and even demonstrating the highest success rate in predicting desorption. The presence of explicit molecular and surface data in the MAD dataset significantly improves the model's ability to describe catalytic reactions. We note that these results are still obtained using inconsistent DFT targets, so further improvement would be expected upon consistent evaluation. 

\begin{table*}[t]
\centering\small
\caption{\label{tab:lambench-neb}
Performance of different universal models on the OC20-NEB task of the LAMBench benchmark.
The mean absolute error (MAE) for reaction energy differences ($\Delta E$) and activation energies ($E_a$) are given in eV. 
The success rates (in \%) for transfer, desorption, and dissociation reactions are calculated as a fraction of the corresponding runs for which the MAE in $E_a$ predictions is less than 0.1 eV. 
The best results per column are shown in bold, and the second-best are underlined.}
\vspace{5pt}
\begin{tabular}{l|cc|ccc}
\toprule
Model & MAE $\Delta E$ ($\downarrow$) & MAE $E_a$ ($\downarrow$) & \% Transfer ($\uparrow$) & \% Desorption ($\uparrow$) & \% Dissociation ($\uparrow$) \\
\midrule
PET-OMAD     & 0.331 & 1.289 & 69.14 & \textbf{98.43} & 66.46 \\
\midrule

PET-MPtrj-C         & 1.257 & 2.131 & 73.14 & 92.91 & 79.75 \\
PET-MPtrj-NC        & 2.030 & 2.874 & \B{80.00} & \U{94.49} & \B{85.44} \\
PET-OAM-NC    & 3.439 & 3.879 & 70.29 & 81.89 & 70.89 \\
PET-OAM-C     & 4.210 & 4.438 & 62.29 & 85.83 & 72.78 \\
\midrule
DPA-3.1-3M       & \B{0.234} & \B{1.203} & 66.86 & 81.10 & 63.92 \\
DPA-2.4-7M       & \U{0.306} & \U{1.271} & 68.57 & 69.29 & 77.85 \\
PET-MAD          & 0.471 & 1.408 & 63.43 & \textbf{98.43} & 77.85 \\
SevenNet-l3i5    & 0.546 & 1.423 & 77.71 & 87.40 & 75.95 \\
MACE-MPA-0       & 0.566 & 1.459 & 72.57 & \U{94.49} & \U{84.81} \\
MACE-MP-0        & 0.580 & 1.468 & 76.57 & 90.55 & \U{84.81} \\
GRACE-2L-OAM     & 0.704 & 1.583 & 65.14 & 90.55 & 72.15 \\
SevenNet-MF-ompa & 1.278 & 2.070 & 66.86 & 92.91 & 68.35 \\
Orb-v3           & 1.470 & 2.298 & 61.71 & 87.40 & 72.15 \\
Orb-v2           & 1.729 & 2.682 & 66.29 & 78.74 & 74.05 \\
MatterSim-v1-5M  & 2.018 & 2.686 & 76.00 & 83.46 & \U{84.81} \\
\bottomrule
\end{tabular}
\end{table*}

The conservative PET-OAM model achieves the best overall accuracy in predicting the shear modulus (Table \ref{tab:lambench-elastic}), and competes closely with other universal models in predicting the bulk modulus. Upon comparing the results of the conservative (PET-OAM-C and PET-MPtrj-C) models against their non-conservative analogs (PET-OAM-NC) and (PET-MPtrj-NC), we clearly see that elastic moduli are very sensitive to the lack of energy conservation. This comes as no surprise, as geometry optimization is part of the protocol for calculating elastic moduli, and this it can fail upon having no relaxation constraints and no rotational averaging, as we show in Sections \ref{ssec:geop-and-phonons} and \ref{sec:phonons}. MAD-trained models (PET-MAD and PET-OMAD) show comparatively worse results due to a difference in the underlying DFT. This observation also indicates that some of the materials properties (like the elastic moduli) can be actually very sensitive to a choice of DFT flavor and settings. 
    
\begin{table*}[t]
\begin{minipage}{0.45 \textwidth}
\centering\small
\caption{\label{tab:lambench-elastic}
Performance of different universal models in predicting the elastic constants from the LAMBench benchmark.
The mean absolute errors (MAE) for shear modulus ($G_\mathrm{VRH}$) and bulk modulus ($K_\mathrm{VRH}$) are given in GPa. 
The best results per column are shown in bold, and the second-best are underlined.}
\vspace{5pt}
\begin{tabular}{l|cc}
\toprule
Model & MAE $G_\mathrm{VRH}$ & MAE $K_\mathrm{VRH}$ \\
\midrule
PET-OAM-C     & \textbf{8.700} & 9.040 \\
\midrule
PET-MPtrj-C         & 12.072         & 16.959 \\
PET-OMAD     & 15.610         & 17.633 \\
PET-OAM-NC    & 23.041         & 23.756 \\
PET-MPtrj-NC        & 41.932         & 20.462 \\
\midrule
GRACE-2L-OAM     & \U{9.138}      & \textbf{7.459} \\
SevenNet-MF-ompa & 9.540          & 9.463 \\
Orb-v3           & 9.749          & \U{7.582} \\
MACE-MPA-0       & 10.270         & 15.026 \\
DPA-3.1-3M       & 10.766         & 10.131 \\
MatterSim-v1-5M  & 12.751         & 14.948 \\
PET-MAD          & 17.325         & 32.559 \\
DPA-2.4-7M       & 17.759         & 16.456 \\
SevenNet-l3i5    & 19.421         & 9.934 \\
MACE-MP-0        & 26.195         & 11.006 \\
Orb-v2           & 66.074         & 44.082 \\
\bottomrule
\end{tabular}
\end{minipage}\ \ \ \ \ \ 
\begin{minipage}{0.45\textwidth}
\centering\small
\caption{\label{tab:lambench-wiggle150}
Performance of different universal models on the Wiggle150 test from the LAMBench benchmark.
The mean absolute error (MAE) and root mean squared error (RMSE) in predicting the energies of the conformers are given in kcal/mol. 
The best results per column are shown in bold, and the second-best are underlined.}
\vspace{5pt}

\begin{tabular}{l|cc}
\toprule
Model & MAE  & RMSE  \\
\midrule
PET-OAM-NC    & \U{6.127} & \U{7.741} \\
\midrule
PET-OAM-C     & 7.219 & 8.166 \\
PET-MP-C         & 8.385 & 9.653 \\
PET-OMAD     & 10.967 & 11.558 \\
PET-MP-NC        & 11.085 & 13.400 \\
\midrule
DPA-3.1-3M       & \textbf{5.669} & \textbf{6.523} \\
Orb-v2           & 6.463 & 8.270 \\
PET-MAD          & 8.798 & 9.818 \\
MatterSim-v1-5M  & 10.730 & 12.450 \\
SevenNet-MF-ompa & 10.970 & 12.800 \\
Orb-v3           & 11.922 & 12.894 \\
GRACE-2L-OAM     & 12.140 & 13.994 \\
SevenNet-l3i5    & 13.881 & 15.098 \\
DPA-2.4-7M       & 14.843 & 15.698 \\
MACE-MPA-0       & 14.915 & 16.677 \\
MACE-MP-0        & 26.597 & 28.513 \\
\bottomrule
\end{tabular}

\end{minipage}
\end{table*}

Finally, our results on the Wiggle150 test show that PET-OAM-NC achieves the best results in predicting the energies of molecular conformers among the trained PET models, and second-best result overall, only falling behind the DPA-3.1-3M model. This can be explained by the presence of explicit molecular data in the DPA-3.1-3M training set. The OAM dataset is designed to primarily describe inorganic materials with no information on molecules. However, the results across the various parts of the LAMBench indicate that PET-OAM models can still afford good predictions on molecular structures.

\section{MADBench evaluation}\label{app:madbench}

MADBench is another minimalistic benchmark that is used to assess the accuracy of universal atomistic models across various domains, ensuring internal consistency of the reference ab initio theory \citep{petmad}. It contains subsets sampled from popular datasets for atomistic machine learning, covering inorganic materials (represented by subsets of the MAD~\citep{mad}, MPtrj~\citep{mptrj-and-chgnet}, matbench-discovery~\citep{matbench}, and Alexandria~\citep{alexandria} datasets), molecules (represented by subsets of the SPICE~\citep{spice} and MD22~\citep{md22} datasets), and catalytic applications (a subset of the Open Catalyst 2020~\citep{oc20} dataset), recomputed with a unified set of DFT settings to obtain internally coherent data on target energies, forces and stresses. 

We evaluate the performance of our proposed models on MADBench and compare it against other model results from Ref.~\citep{petmad}. Here, we use the same set of the preliminary conservative and non-conservative models, mentioned above in Sec.~\ref{app:lambench}.
The results are presented in Table~\ref{tab:madbench}. In almost all cases, the PET-OMAD model demonstrates the best overall accuracy in predicting both energies and forces, while falling behind PET-MPtrj and PET-OAM only on the MPtrj and Alexandria subsets, which have considerable overlap with the noted subsets in the training data. A similar effect is observed in the Orb-v2 results for Alexandria. The best accuracy in predicting forces from this subset is likely due to overfitting on the Alexandria data, which is part of this model's training set. Upon comparing the PET-OMAD results against PET-MAD, one can see the combined effect of pre-training on the OMat24 dataset and using the updated architecture: PET-MAD errors are cut almost by half on the majority of the subsets.

\begin{table*}[t]
\centering\small
\caption{\label{tab:madbench}
Evaluation of various universal MLIPs on the MADBench benchmark from Ref.~\citep{petmad}. Performance of the PET-OAM, PET-MPtrj and PET-OMAD models is computed in this work, and other results are reproduced from the reference paper~\citep{petmad}. For each subset and model, mean absolute errors are reported for raw energy (E) and force (F) predictions in meV/atom and meV/\AA, respectively. We do not show force
errors for the matbench-discovery subset, since forces are not available in the reference data.}
\vspace{3pt}
\begin{tabular}{l|cc|cc|c|cc|cc|cc|cc}
\toprule
\multicolumn{1}{c|}{Model}
  & \multicolumn{2}{c|}{MAD}
  & \multicolumn{2}{c|}{MPtrj}
  & Matbench
  & \multicolumn{2}{c|}{Alexandria}
  & \multicolumn{2}{c|}{OC2020}
  & \multicolumn{2}{c|}{SPICE}
  & \multicolumn{2}{c}{MD22} \\
\midrule
 & E & F & E & F & E & E & F & E & F & E & F & E & F \\
\midrule
PET-OMAD & \B{6.0} & \B{29.2}	& 12.1 & 36.0 & \B{10.6} & 26.3 & 30.4	& \B{8.0} &  \B{51.6}& 	\B{2.4} & \B{44.6}	& \U{3.4} & \B{53.4} \\ 
\midrule
PET-OAM-C     & 43.1 & \U{58.3} & 10.1 & 13.5 & 37.1 & \U{12.6} & 21.3 & 25.8 & \U{67.7}  & 13.2 & \U{53.6} & 10.9 & \U{59.3} \\
PET-OAM-NC    & 46.3 & 71.4 & \U{4.0}  & 11.9 & 38.6 & \B{11.4} & \U{12.5} & 20.0 & 76.5  & 13.7 & 75.6 & 13.4 & 93.0 \\
PET-MPtrj-C      & 55.8 & 101.3 & 5.5  & \U{9.1}  & 41.8 & 36.5 & 51.7 & 42.4 & 117.3 & 13.0 & 81.8 & 7.4  & 101.0 \\
PET-MPtrj-NC     & 91.3 & 102.2 & \B{2.9}  & \B{7.8}  & 48.7 & 44.6 & 47.7 & 34.7 & 119.1 & 34.7 & 103.1 & 19.4 & 122.8 \\
\midrule
PET-MAD       & \U{17.6} & 65.0 & 22.3 & 77.6 & \U{31.3} & 49.0 & 65.3 & \U{18.3} & 114.5 & \U{3.7}  & 59.4 & \B{1.9}  & 65.6 \\
MACE-MP-0-L   & 81.6 & 181.5 & 15.1 & 50.8 & 58.5 & 65.4 & 79.5 & 82.4 & 169.6 & 10.6 & 166.8 & 9.4  & 182.9 \\
MatterSim-5M  & 47.3 & 133.7 & 21.3 & 61.4 & 38.2 & 21.2 & 39.9 & 31.5 & 119.2 & 21.3 & 145.6 & 28.6 & 160.4 \\
Orb-v2        & 52.9 & 96.2 & 5.6  & 21.9 & 37.9 & 13.2 & \B{10.5} & 19.8 & 99.3  & 59.0 & 140.8 & 174.3& 220.7 \\
SevenNet-l3i5 & 82.1 & 173.5 & 9.8  & 25.5 & 47.5 & 47.6 & 70.3 & 45.7 & 162.7 & 11.3 & 139.1 & 11.1 & 146.2 \\
\bottomrule
\end{tabular}
\end{table*}

\clearpage
\section{SPICE evaluation}\label{app:spice}

\begin{figure}[ht]
    \centering
    \includegraphics[width=\linewidth]{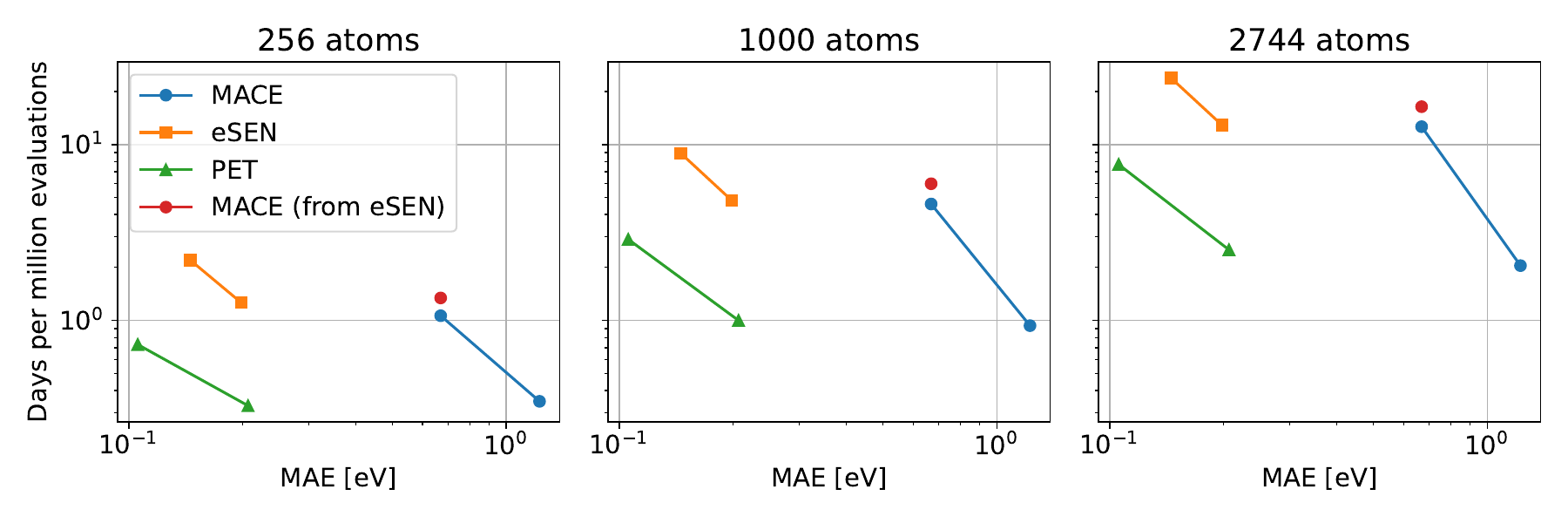}
    \caption{Accuracy-speed Pareto front for models trained on the SPICE dataset, at varying number of atoms. Timings corresponding to MACE evaluations in Ref.~\citep{esen} do not differ significantly from those presented in this work and might stem from different hardware power settings.}
    \label{fig:spice-appendix}
\end{figure}

In order to obtain timings for the MACE, eSEN and PET architectures, we perform inference with the same set-up described in Ref.\citep{esen}, using the same hardware (a single Nvidia A100 80 GB GPU). The timings we present in this work are evaluated by us for PET and MACE, while the eSEN timings are taken from Ref.\citep{esen}. In Fig.~\ref{fig:spice-appendix} we further include the MACE-L timings reported in Ref.~\citep{esen}, showing that any potential differences due to the evaluation setup do not affect the results significantly (we suspect them to originate from different power GPU power settings on different clusters). Similarly, although we make use of torchscript compilation for PET evaluations (while MACE and eSEN do not), the differences we observed compared to eager-mode Pytorch~\citep{pytorch} evaluations are minimal.

For completeness, we report accuracies of all evaluated models in Tab.~\ref{tab:spice-appendix}. MACE and eSEN accuracies are taken from Refs.~\citep{mace-off} and~\citep{esen}, respectively.

\begin{table}[ht]
\caption{Test set accuracies for models trained on the dataset presented in Ref.~\citep{mace-off}. Energy MAEs are shown in units of meV per atom; force MAEs are shown in units of meV/\AA. For each metric, the best model is highlighted in bold and the second best is underlined.}
\label{tab:spice-appendix}
\begin{center}
\begin{tabular}{l|cc|cc|cc|cc|cc|cc}
\toprule
\multicolumn{1}{c}{Subset} & \multicolumn{2}{c}{MACE-S} & \multicolumn{2}{c}{MACE-L} & \multicolumn{2}{c}{eSEN-S} & \multicolumn{2}{c}{eSEN-L} & \multicolumn{2}{c}{PET-S} & \multicolumn{2}{c}{PET-L} \\
\midrule
 & E & F & E & F & E & F & E & F & E & F & E & F \\
\midrule
PubChem               & 1.41 & 35.68 & 0.88 & 14.75 & 0.22 &  6.10 & \U{0.15} & \U{4.21} & 0.19 &  6.79 & \B{0.09} & \B{3.53} \\
DES370K Monomers      & 1.04 & 17.63 & 0.59 &  6.58 & 0.17 &  1.85 & \U{0.13} & \U{1.24} & 0.19 &  2.23 & \B{0.10} & \B{1.00} \\
DES370K Dimers        & 0.98 & 16.31 & 0.54 &  6.62 & 0.20 &  2.77 & \U{0.15} & \U{2.12} & 0.18 &  2.17 & \B{0.12} & \B{1.20} \\
Dipeptides            & 0.84 & 25.07 & 0.42 & 10.19 & 0.10 &  3.04 & \U{0.07} & \U{2.00} & 0.11 &  3.58 & \B{0.05} & \B{1.55} \\
Solvated Amino Acids  & 1.60 & 38.56 & 0.98 & 19.43 & 0.30 &  5.76 & \U{0.25} & \B{3.68} & 0.36 &  8.78 & \B{0.17} & \U{4.37} \\
Water                 & 1.67 & 28.53 & 0.83 & 13.57 & 0.24 &  3.88 & \U{0.15} & \B{2.50} & 0.27 &  6.11 & \B{0.13} & \U{3.05} \\
QMugs                 & 1.03 & 41.45 & 0.45 & 16.93 & 0.16 &  5.70 & \U{0.12} & \U{3.78} & 0.15 &  6.71 & \B{0.08} & \B{2.91} \\
\bottomrule
\end{tabular}
\end{center}
\end{table}

\section{The Open Molecules Benchmark}\label{app:omol}

We trained additional models on the OMol-1 dataset~\cite{omol} containing 140M systems.  To accommodate the varying charge states and spin multiplicities present in the dataset, we introduced a system-conditioning embedding into the PET architecture that closely follows the embeddings introduced in Ref.~\cite{omol}.
Separate embeddings for charge and spin states, each matching the dimensionality of the node features, are projected onto the node dimensions via an additional linear layer. These embeddings are then added to the existing node embeddings at each message-passing step. We initialize the projection layer to zero to stabilize training and to enable seamless combination of the system conditioning with already-pretrained models. During training, energies were pre-normalized by subtracting atomic energies using the neutral atomic reference values provided with the dataset, followed by additional linear fit. To enable efficient training despite the varying molecule sizes in OMol, we use max-atom batching rather than a fixed system number. Similar to the implementation in the \texttt{fairchem} package, we greedily fill batches until the next system would exceed the maximum number of atoms per batch, then start a new batch. Batches are created once, and their order is shuffled each epoch.
The small, medium, and large models were trained for 15 epochs with a batch size of 64,000 atoms, corresponding to an average of 1,454 systems per batch. The extra-small model was trained for 10 epochs. The medium model was adapted to the domain by adding one additional message-passing layer; otherwise, the model hyperparameters correspond to those used in the OMat training exercises.

The conservative models were fine-tuned for 7 epochs with a batch size of 128,000 atoms and the learning rate of 0.0002 was reduced to 0.0001 for the fine-tuning. We note that all models (and especially the conservatives models) would benefit from additional training, but we were constrained by available GPU resources and we plan to publish improved models in the future. Results for all models evaluated on the OMol validation splits without rotational symmetrization can be found in Table~\ref{tab:errors}. Metrics for the large conservative model with rotational symmetrization on the full FairChem benchmark have been published on the official leaderboard. The models (as well as newer versions) will be available as part of the \texttt{upet} repository.

 \begin{table}[htbp]
\centering
\caption{Energy (eV) and force MAE (eV/$\AA$) for conservative and non-conservative PET models on the public OMOL validation set.}
\label{tab:errors}
\vspace{0.5em}
\textbf{(a) Conservative models}
\vspace{0.3em}

\begin{tabular}{lccc}
\hline
Model    & Energy MAE & Energy MAE/atom & Force MAE \\
\hline
PET-S  & 0.438 & 0.0040 & 0.0160 \\
PET-M & 0.153 & 0.0014 & 0.0078 \\
PET-L  & 0.082 & 0.00093 & 0.0071 \\
\hline
\end{tabular}

\vspace{1em}
\textbf{(b) Non-conservative models}
\vspace{0.3em}

\begin{tabular}{lccc}
\hline
Model & Energy MAE & Energy MAE/atom & Force MAE \\
\hline
PET-XS & 0.191 & 0.0024 & 0.0212 \\
PET-S  & 0.166 & 0.0021 & 0.0197 \\
PET-M  & 0.0798 & 0.0010 & 0.0086 \\
PET-L  & 0.0707 & 0.00095 & 0.0072 \\
\hline
\end{tabular}
\end{table}

\section{Details on geometry optimization and phonon calculations}\label{sec:phonons}

\begin{figure}[tb]
    \centering
    \includegraphics[width=\linewidth]{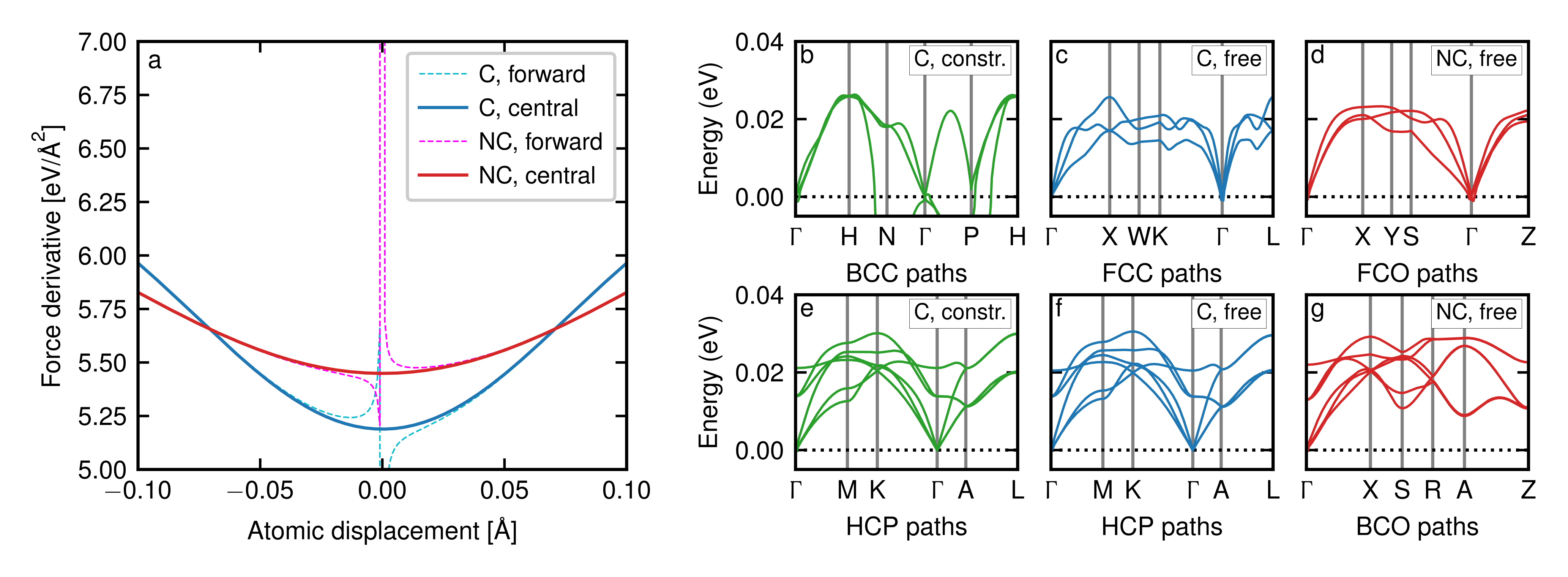}
    \caption{(a) Finite-difference force derivative on one HCP Ti atom as a function of its displacement from symmetry-constrained equilibrium. (b-g) Phonon bands plotted along high-symmetry lines in the BZ corresponding to the relaxed structures.}
    \label{fig:forces_and_phonons-appendix}
\end{figure}

Geometry optimizations are performed with ASE~\citep{ase-paper} version 3.26.0 using metatomic~\citep{metatensor} calculators. Constrained optimizations use the FixSymmetry function and freezing the off-diagonal strain degrees of freedom in FrechetCellFilter. The convergence threshold for forces and (appropriately scaled) stress components is set to $10^{-5}$ eV/\AA.

Phonon bands are computed with ASE's phonons module. The displacement used for finite-difference force constants is 0.03 \AA. All calculations of interatomic force constants are done using central finite differences (rather than forward differences, as commonly done from first principles~\cite{Seko2015} and with equivariant models in matbench-discovery), to prevent numerical instabilities at small displacements caused by residual forces in the symmetry-constrained minimum (see Figure~\ref{fig:forces_and_phonons-appendix}a).

In Figure~\ref{fig:forces_and_phonons-appendix}(b--g) we plot again the phonon spectra from Figure~\ref{fig:geomopt_and_phonons}, this time along the high-symmetry lines of the BZ corresponding to the Bravais lattice of each relaxed structures. 
The lattices are identified using spglib~\citep{spglib} with a tolerance (``symprec'') of 0.01. With this setting, unconstrained relaxations using the C model return FCC and HCP structures, while NC relaxations yield face-centered orthorombic (FCO) when starting from BCC and base-centered orthorombic (BCO) when starting from HCP.
Relaxing the tolerance to 0.1 assigns the former as body-centered tetragonal and the latter as HCP, whereas tightening it to 0.001 reduces all unconstrained relaxations to triclinic $\mathrm{P\overline{1}}$.

\end{document}